# Study of the structural and electronic properties of the Heusler Co2FeGe alloy by DFT approach

A. JAMRAOUI [1], Y. SELMANI [1], A. JABAR [2,3], L. BAHMAD [1]

[1] Laboratory of Condensed Matter and Interdisciplinary Sciences (LaMCScI), Faculty of Sciences, Mohammed V University in Rabat, Av. Ibn Batouta, B. P. 1014 Rabat, Morocco.

[2] Laboratory of Mechanics and High Energy Physics, Department of Physics, Faculty of Sciences Aïn Chock, University Hassan II, P.O. Box 5366 Maarif Casablanca 20100, Morocco

[3] Laboratory of Condensed Matter and Interdisciplinary Sciences (LaMCScI), Faculty of Sciences, Mohammed V University in Rabat, Av. Ibn Batouta, B. P. 1014 Rabat, Morocco

*Corresponding author: l.bahmad@um5r.ac.ma (L.B.)

Abstract:

In this work we reported the structural and electronic properties of the Heusler compound Co2FeGe using the AKAI-KKR code under the GGA approximation. We established that this material presents not only magnetic character but also has a metallic behavior. Our calculations have been conducted using the DFT method in the framework of the AKAI-KKR code. This study enabled us to define certain characteristics and initial parameters for creating a model of the system. The method used allowed us to apply fundamental concepts to the studied system in the form of modeling. The main results, of the studied Heusler compound Co2FeGe are: i) this material is magnetic; ii) The band structure of the material predicts a metallic character; iii) the origin of magnetism comes mainly from the transition metals Co and Fe atoms. These results, assure that the studied quaternary Heusler Co2FeGe stands for a strong candidate for different spintronics applications.

# 1. Introduction:

Both technological and industrial developments depend heavily on the search for new materials and alloys from the periodic table of elements based on the natural law which states that the combination of two different materials does not present a combination of their properties but rather gives rise to new characteristics specific to the alloy [1].

Since the prediction of the semi-metallicity of the Heusler alloy NiMnSb by De Groot et al. in 1983, the scientific interest in Heusler alloys has been renewed and these materials have started to have both theoretical and experimental interests and especially in spintronics. Moving from electronics based on the control of charge currents, negatively charged electrons and positively charged holes, to a new concept called spin electronics, or spintronics, an emerging field that plans to use the spin of electrons as an additional degree of freedom to generate new and modern functionalities. Several Heusler alloys have then been predicted by ab-initio calculations [2].

Meanwhile the discovery of the ferromagnetic half-metallicity of Heusler alloys, they have become a field of research interest for spin electronics applications [3]. The term Heusler alloys is given to a group of compounds that contains approximately 3000 identified compounds. These remarkable compounds were first discovered by Fritz Heusler in 1903 while working on the ferromagnetism of the Cu2MnAl alloy [4]. Due to a wide range of properties, these compounds can behave as semimetals, semiconductors, superconductors, and many others.

In rare cases, element *Y* can be a rare earth element or an alkaline earth metal. Generally, the doubled atom *X* is at the beginning of the formula, and the atom *Z* from group III, IV or V at the end, such as Co2MnSi [5, 6]. Exceptions can be found where the order of ranking follows the electronegativity scale, for example the element LiCu2Sb [7].

A semimetal is a material that behaves like a metal in one spin direction (either "spin-up" or "spin-down") and like an electrical insulator or semiconductor in the opposite direction [8]. This feature is called semi-metallic ferromagnetism [9, 10]. Although semimetals are ferromagnetic, most ferromagnetic substances are not semimetals.

The term half-metallicity was first introduced by Groot et al [11] in the early 1980s who were interested in calculating the band structure for the half-Heusler alloy NiMnSb [12], only electrons of a given spin orientation ("up" or "down") have metallic character, while electrons of the other spin orientation have insulating or semiconductor behavior.

Conventional ferromagnetic materials exhibit an electronic density of states (N(EF)) at the Fermi level for the majority spin (N↑(EF)) and minority spin (N↓(EF)) electrons, represented in Figure I.7. The definition of spin polarization (P), is the measure of spin asymmetry, it is given by the following relation [13].

An interesting study was carried out by Mouatassime et al. [14], where the structural optimization showed the semi-metallic character of the Full-Heusler Co2FeGe compound through the use of the GGA+U approximation method. Other works allowed to give good results to highlight the semi-metallic gap of the material [15]. After optimizing the structure, they implemented the Coulomb repulsion potential in the local density approximation, and the comparison of the results obtained with LSDA, and LSDA+U. As well as the impact of the Coulomb repulsion to show the semi-metallicity of the full-Heusler Co2FeGe.

The Co2FeGe sample was prepared by arc melting method, first the sample is returned to atmospheric pressure under argon to avoid any form of oxidation, then the pure elements (Co-99.99%, Fe-99.7%, Ge-99.99%) were placed in a chamber containing cavities of the water-cooled copper anode of an induction furnace and heated up to 1000 °C because this temperature is large compared to the melting temperature of the prepared element, then the element is removed and cooled, these operations are repeated 4 or 5 times, magnetic measurements were carried out with MPMS (Magnetic Cleanliness Measurement System) and the lattice parameters by X-ray diffraction [16].

The Co2FeGe material is a ferromagnetic material with an experimental total magnetic moment of $\mu = 5.54 \, \mu B$ and a transition temperature of 1060 K [17].

In this work, we present the calculation results of the structural stability and electronic properties of the Co2FeGe compound. By the AKAI-KKR code, using the GGA approximation [18], associated with the density functional theory (DFT) [19].

## 2. Results and discussion

### a) Structural properties

We have visualized the crystal structure using the Xcrysden software, which shows that the Heusler Co2FeGe is a face-centered cubic (FCC) structure for Iron (Fe). The cobalt (Co) atoms occupy all 8 tetrahedral sites, while the germanium (Ge) atoms occupy all 12 octahedral sites of the FCC unit cell. The positions (coordinates) of the atoms in the unit cell of the Co2FeGe compound, as seen in Fig. 1, are:

- Fe: (0, 0, 0).
- Co:(1/4,1/4,1/4), (1/4,3/4,1/4), (3/4,1/4,1/4), (3/4,3/4,1/4), (1/4,1/4,3/4), (1/4,3/4,3/4), (3/4,1/4,3/4), (3/4,3/4,3/4).
- Ge:(1/2, 1/2, 1/2),(1/2,1/2,0),(1/2,0,1/2),(0,1/2,1/2).

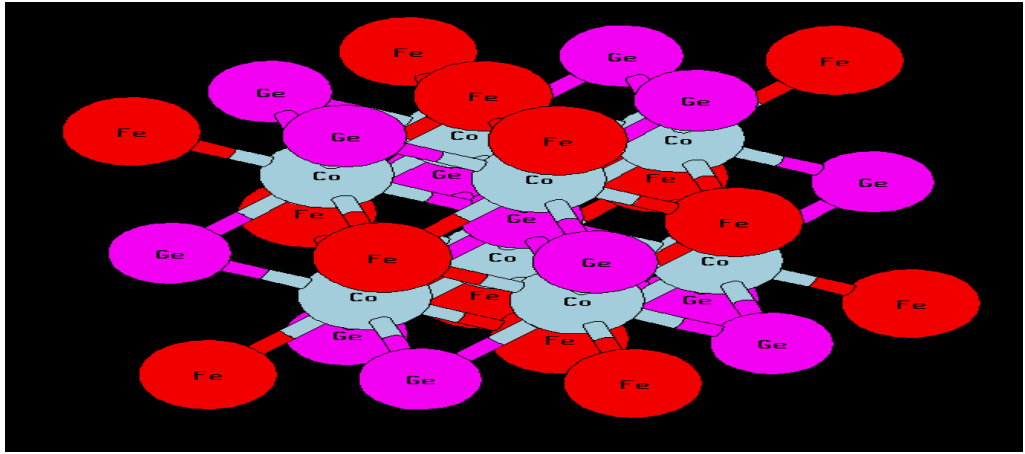

*Figure 1: Crystal structure of the Co2FeGe compound by Xcrysden.*

Figure 1 displays the crystal structure of the Heusler Co2FeGe for a unit cell, composed of three different types of atoms, represented by different colors:

- Red spheres: Labeled "Fe", represent Iron atoms.

- Purple spheres: Labeled "Ge", stand for Germanium atoms.

- Light blue spheres: Labeled "Co", are Cobalt atoms.

The arrangement shows a repeating pattern where the atoms are bonded together. This type of image is commonly used in materials science and chemistry to visualize the atomic structure of intermetallic compounds or alloys.

The specific arrangement suggests a complex crystal structure, and without further information, it's difficult to definitively identify the crystal system or space group. However, it clearly depicts a compound containing Iron, Germanium, and Cobalt.

To determine the various ground state properties, it is necessary to optimize the total energy for the system under study. This step allows us to predict the most stable phase in which the material crystallizes. Once equilibrium is reached, we can access various physical properties (electronic, magnetic, etc.).

Structural optimization is performed by minimizing the total energy as a function of volume (V). The optimization cycle is repeated until convergence is achieved. The equilibrium bulk modulus is evaluated by fitting the obtained curve of the total energy variation as a function of volume.

The curve shown in Figure 2 indicates that the energy has a minimum at an optimum lattice parameter. This optimum lattice parameter corresponds to the lattice parameter optimized at a temperature of T=0 K. This figure summarizes the relationship between Energy and Volume (Å³). The curve shows a distinct parabolic or nearly parabolic shape, opening upwards. There is a clear minimum point corresponding to the equilibrium volume and the minimum appearing to be around 190-200 Å³. The corresponding minimum energy is approximately -12274.5 Ry.

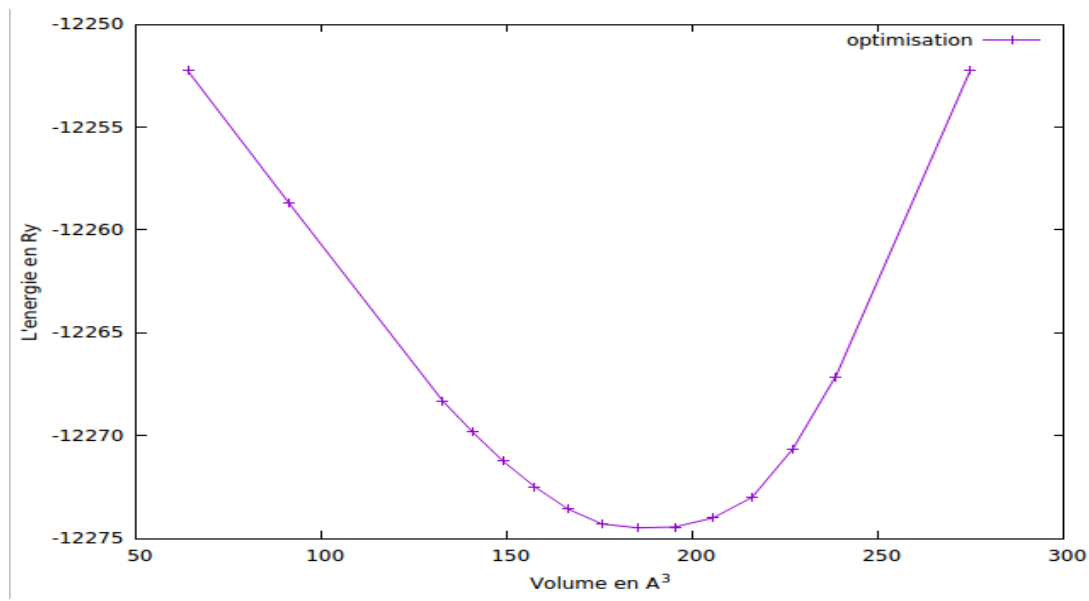

*Figure 2: Variation of the total energy as a function of volume for the $Co_2FeGe$ compound.*

Table 2 summarizes the equilibrium and experimental lattice parameter a(Å), as well as the total energy (Ry) for the Co2FeGe compound.

|  | Lattice parameter a(Å) of Co2FeGe | Energy (Ry) of Co2FeGe |
|---|---|---|
| Expérimental | 5.738 [20] | --- |
| GGA | 5.700 | -12274.47513 |

*Table 2: Lattice parameter a(Å) at equilibrium and experimental, as well as the total energy (Ry) for the Co2FeGe compound.*

b) Electronic Properties

The electronic properties of the studied compound (band structure and density of state) depend mainly on the distribution of electrons in the valence and conduction bands, as well as on the gap value. These parameters are calculated for the Co2FeGe compound in their equilibrium state with the optimal lattice parameter.

- Band Structure

The most significant description of the energy surfaces offered to electrons is performed in reciprocal space or k-wave vector space. This description is generally simplified by considering the variations of energy E as a function of k along the directions of highest symmetry in this space. In these directions, and by limiting ourselves to the first Brillouin zone.

Conduction and valence bands are multiple, but electronic transport properties depend primarily on the structure of the lowest conduction band (BC) and the highest valence band (BV).

Using the AKAI-KKR code, the electronic band structures of the compound Co2FeGe were calculated along the directions of high symmetry, using the GGA approximation. The results are presented in Figure 3 for spin-up (left) and spin-down (right).

The Fig. 3(left) provided the band structure diagram of spin-up for the Co2FeGe compound. The plots describe the fundamental concepts in solid-state physics for the allowed energy levels for electrons within a crystalline material. The different high-symmetry points and lines in the Brillouin zone are labeled with Greek or Latin letters (e.g., "G", "S") indicating specific directions in the momentum space. The vertical lines indicate the boundaries between these different symmetry directions. The energy is "relative to Fermi energy" meaning that the Fermi energy is set to 0

Ry (at y=0). This energy level at which, for the absolute zero temperature, all states below it, are occupied by electrons, and all states above it, are empty. For spin-up states, there is a clear energy gap around the Fermi level (0 Ry). The top of the valence band appears to be just below 0 Ry, and the bottom of the conduction band starts above 0 Ry. This indicates that the material is a semiconductor for this channel spins. To determine if the material has a direct or indirect band gap, one would need to compare the momentum (k-point) of the valence band maximum with that one of the conduction band minima. If they occur at the same k-point, it's a direct gap; otherwise, it's an indirect gap. From our case, the valence band maximum is located at a different k-point than the conduction band minimum. This situation suggested an indirect gap, but a precise determination would require closer inspection of the specific k-points.

Regarding the spin-down channel of the studied compound, our results are provided in Fig. 3(right). This figure presents different high-symmetry points and lines in the Brillouin zone (reciprocal space) of the crystal lattice. These points indicate specific directions in the momentum space. The vertical lines specify the boundaries between these different symmetry directions. From this plot, there is a clear energy gap around the Fermi level (0 Ry). This confirms that the material is semiconductor. The size of this gap is crucial for determining the material's electrical conductivity. From this figure, it is clear that the valence band maximum is situated at a different k-point than the conduction band minimum. This suggests an indirect gap. Also, this band structure diagram provides critical insights into the electronic properties of the material for the spin-up channel.

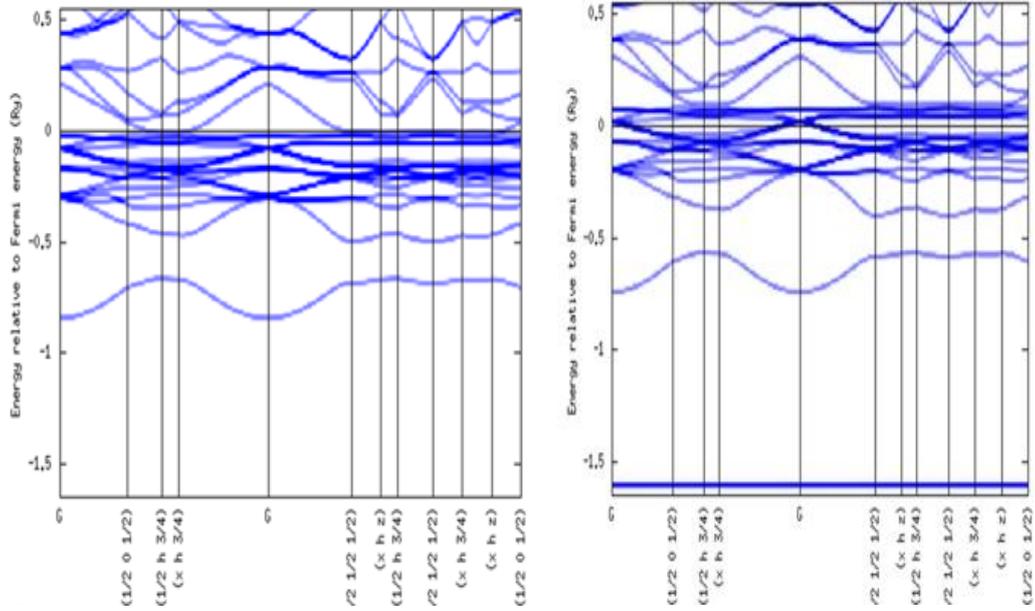

*Figure 3: Band structure of spin-up (left) and spin-down (right) of the compound Co2FeGe by the GGA approximation.*

The compound Co2FeGe is characterized by an overlap of the valence and conduction bands. The most important observation is the presence of electronic states at the Fermi level in the band structure of both spin-up and spin-down electrons. This means that the material studied is a compound with metallic character for both spin orientations.

- Electronic Density of states

In solid-state physics, the density of states (DOS) quantifies the number of electronic states with a given energy in the material under consideration. To determine the nature of the electronic band structure, we also calculated the total and partial densities of states. They thus allow us to determine the nature of the chemical bonds in a material and consequently the charge transfer between orbitals and atoms. The results of our calculations are presented in Figures 4 and 5.

Figure 4 presents the partial density of states of majority and minority spins of for Fe, Co and Ga, for the studied Co2FeGe compound. This figure has been plotted in the frame work of the GGA approximation.

The spin-polarized partial density of states (pDOS), specifically focusing on the d-orbitals, is illustrated in Figs. 4(Fe), 4(Co) and 4(Ga), for the constituent elements Fe, Co and Ga, respectively. The energy value 0 Ry corresponds to the Fermi level. The Y-axis stands for "partial Density of States," which measures the number of available electronic states per unit energy range. A higher DOS value at a particular energy indicates more electronic states are available at that energy.

The most striking feature is the significant difference between the spin-up and spin-down DOS curves. This asymmetry is a clear indication of the ferromagnetism coming from the Iron element. The same behavior is seen in Fig. 4(Co) for Co element. While the spin-up and spin-down DOS symmetrical Ga orbitals, see Fig. 4(Ga), indicates that the Ga element is non-magnetic. For the Fe element, a large peak just below the Fermi level (around -0.05 Ry to 0 Ry) indicates a high density of available spin-up d-states near the Fermi level corresponding to the "majority" spin electrons. Regarding the Spin-down of the Fe element, a prominent peak shifted to higher energies, crosses and extends above the Fermi level (around 0.05 Ry to 0.15 Ry). This implies a lower density of spin-down states occupied below the Fermi level compared to spin-up states, and a higher density of unoccupied spin-down states just above the Fermi level.

At the Fermi level (0 Ry), there is a relatively high density of spin-up states and a very low density of spin-down states. This difference at the Fermi level is crucial for understanding metallic ferromagnetism. The presence of states at the Fermi level means that the corresponding element (Fe or Co) is a metal. The distinct shift in energy between the spin-up and spin-down d-bands is due to the exchange splitting. This is an effect arising from the strong electron-electron interactions (specifically, the exchange interaction) which causes electrons with parallel spins to occupy different energy levels than electrons with anti-parallel spins. This splitting leads to a net magnetic moment.

While this is a DOS plot and not a band structure, the peaks correspond to high densities of states, which in turn relate to flatter bands in the electronic band structure. The sharp peaks near the Fermi level suggest narrow d-bands, characteristic of transition metals.

In summary, this Fig. 4 illustrates the ferromagnetic nature of the studied material, showing how the spin-up and spin-down d-electron states are energetically split, leading to a net magnetic moment.

Figure 5 showed the majority and minority spins of partial density of states of Fe, Co and Ga for the compound Co2FeGe with the approximation GGA. This figure supports the fact that the sutdied compoud exhibits a metallic character. In fact, one can see that the Iron is magnetic and metallic and the same thing for Cobalt. On the other hand Germanium is non-magnetic since its density of states for the majority and minority spins is symmetrical.

The total and partial Density of States of the compound Co2FeGe for the majority and minority spins with the GGA approximation confirms our previous results. In particular, the magnetization of the studied compound Co2FeGe, comes mainly from Iron and Cobalt atoms, see Fig. 5. As the total density of states of the Co2FeGe compound for the majority and minority spins is not symmetric, therefore this material exhibits a magnetic character.

In connection with Fig. 4, the Fig. 5 shows the significant asymmetry between the spin-up (positive DOS) and spin-down (negative DOS) total density of states for Co2FeGe clearly indicates that this material is ferromagnetic. The presence of a substantial DOS at the Fermi level (0 Ry) for both spin-up and spin-down channels, especially for the spin-up channel, suggests that Co2FeGe assures the metallic character of this material.

In addition, such compounds are often studied for their magnetic properties, especially for their potential as half-metallic behavior. Consequently, the large peaks in the TDOS are predominantly contributed by the d-orbitals of both Co and Fe atoms. This is confirmed by the strong overlap of their partial PDOS curves shown in Fig. 4. This is typical for transition metal-based compounds where d-electrons play a crucial role in electronic and magnetic properties.

On the other hand, the Fe (d) partial PDOS (Fig. 4 Fe) shows a very significant contribution to both the spin-up and spin-down DOS, in particular around the Fermi level. This plot exhibits a strong peak for spin-up just below the Fermi level and also a strong peak for spin-down just above the Fermi level.

Regarding the Co (d) partial PDOS (Fig. 4 Co), this element contributes substantially in the total DOS with its spin-up peak just below the Fermi level. This orbital overlaps with the Fe contribution one, see Fig. 5. In fact, the corresponding spin-down contribution is also significant, with a peak slightly above the Fermi level.

Concerning the contribution of the Ge (d) partial orbital, as expected, the corresponding plot shows a very small contribution near the Fermi level, indicating that its d-electrons are not primarily responsible for the electronic states in this energy range. Its main contribution would likely come from its p-orbitals, which are not explicitly plotted as partial PDOS.

The distinct energy separation between the spin-up and spin-down in the d-states of Co and Fe, points to strong exchange splitting and hybridization between the d-orbitals of Co and Fe atoms.

This hybridization leads to the formation of molecular orbitals that contribute to the overall electronic structure and magnetic moment.

In summary, a perfectly half-metallic ferromagnet behavior would have a metallic character for one spin channel and a semiconductor or insulating character for the other spin channel. In fact, the spin-up channel clearly shows metallic behavior with a significant DOS at 0 Ry. For the spin-down channel, while the DOS is significantly lower than for spin-up, it is not null at 0 Ry. Therefore, based on the Fig. 5 plot, Co2FeGe appears to be a metallic ferromagnet, but not a perfect half-metal. There might be some small minority-spin states at the Fermi level, preventing it from being perfectly half-metallic.

In conclusion, the DOS plot of Co2FeGe provides a detailed picture of the electronic and magnetic properties, highlighting its ferromagnetic nature, the metallic character, and the significant roles of Co and Fe d-electrons in determining these properties.

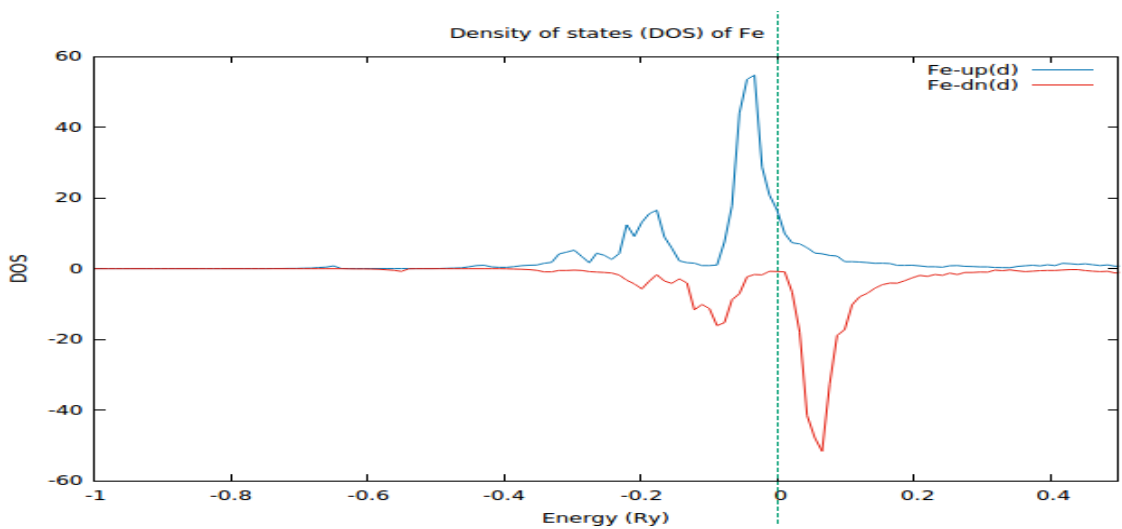

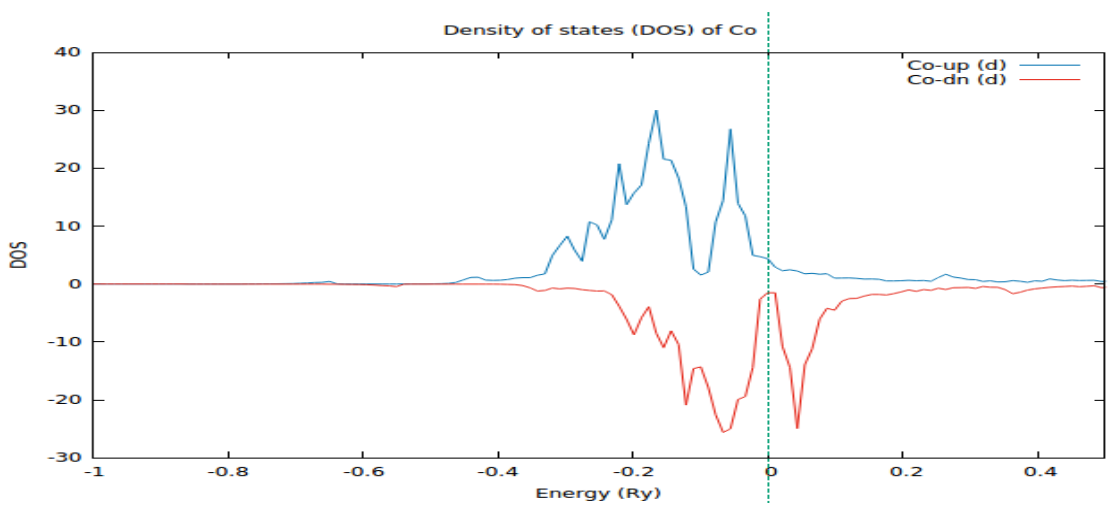

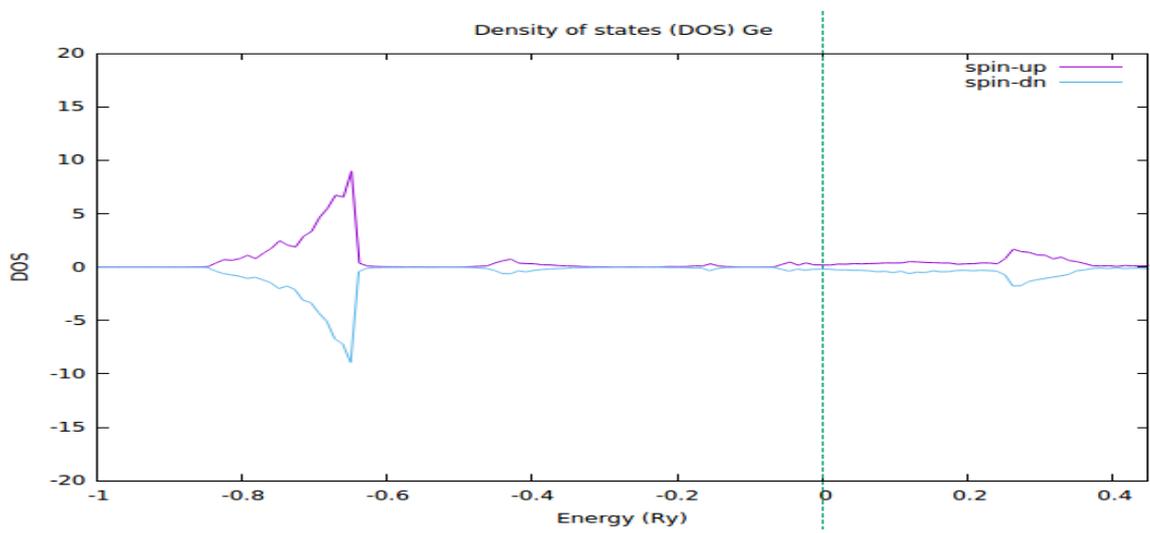

*Figure 4: Partial density of states of Fe, Co and Ga for the compound Co2FeGe showing majority and minority spins with the approximation GGA.*

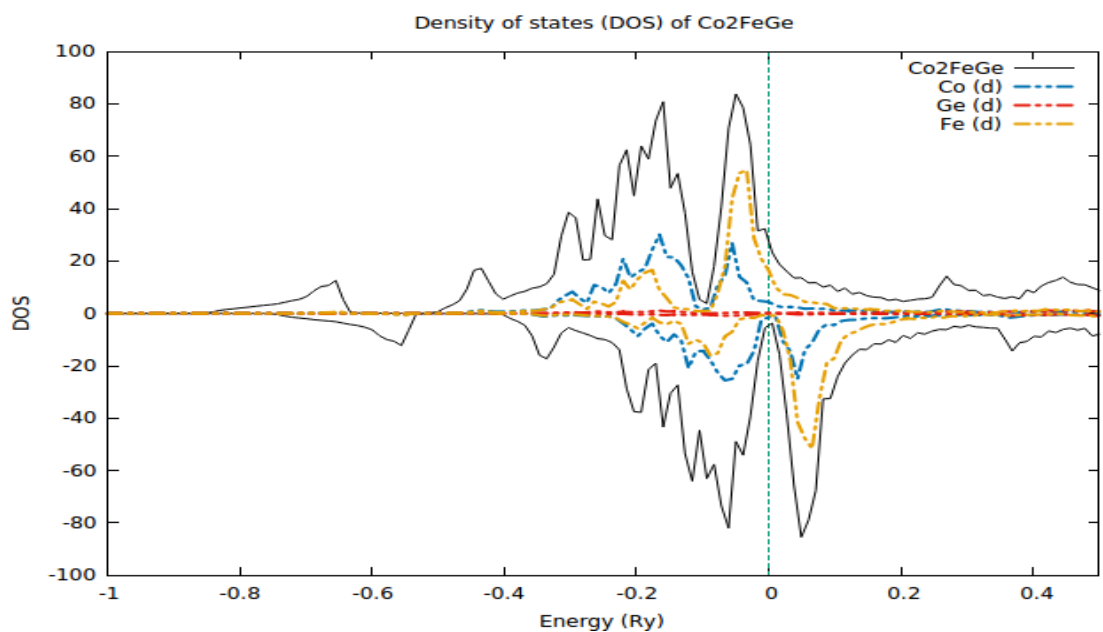

*Figure 5: Total and partial density of states of the compound Co2FeGe for majority and minority spins with the GGA approximation.*

## 3. Conclusion

We have studied the structural and electronic properties of the compound Co2FeGe by the AKAI-KKR code with the GGA approximation, and we have shown that this material is magnetic and has a metallic character. We modeled the Co2FeGe compound using the DFT method with the AKAI-KKR code. The proposed model allowed us to study its structural and electronic properties.

The properties of materials are constantly being explored in various aspects of our daily lives. This makes learning and delving deeper into the field of materials, along with their simulation and modeling, a necessity.

This study enabled us to define certain characteristics and initial parameters for creating a model of the system. The method used allowed us to apply fundamental concepts to the studied system in the form of modeling using the Ab-initio calculations with AKAI code, facilitating the determination of magnetic quantities and the study of the system at T=0K.

The main results are summarized as follows:

---

- We concluded that the compound is magnetic.
- The band structure of the material predicts a metallic character.
- The origin of magnetism in this Heusler material primarily comes from the transition metals Co and Fe.

---

Based on these results, it is evident that the studied quaternary Heusler compound is a strong candidate for future applications in spintronics.

## Funding

The authors declare that no funding was received for this research.

## Conflicts of interest

The authors declare that there is no conflict of interest.

**Appendix:** Input preparation for the Co2FeGe Heusler:

We create our AKAI KKR input using the Xband software. We start by choosing the system creation method. In our case, we opted to create the crystal system using the space group of the Co2FeGe compound, which is Fm-3m[225].

Hence, we insert the atomic positions of Cobalt (Co), Germanium (Ge), and Iron (Fe) that form the unit cell. The rest of the lattice is then obtained by translating these coordinates and by symmetry. And then we input the lattice parameters given in the literature: a=b=c = 4.057 Angstrom, and convert them into atomic units (a.u.), see Fig. 1.

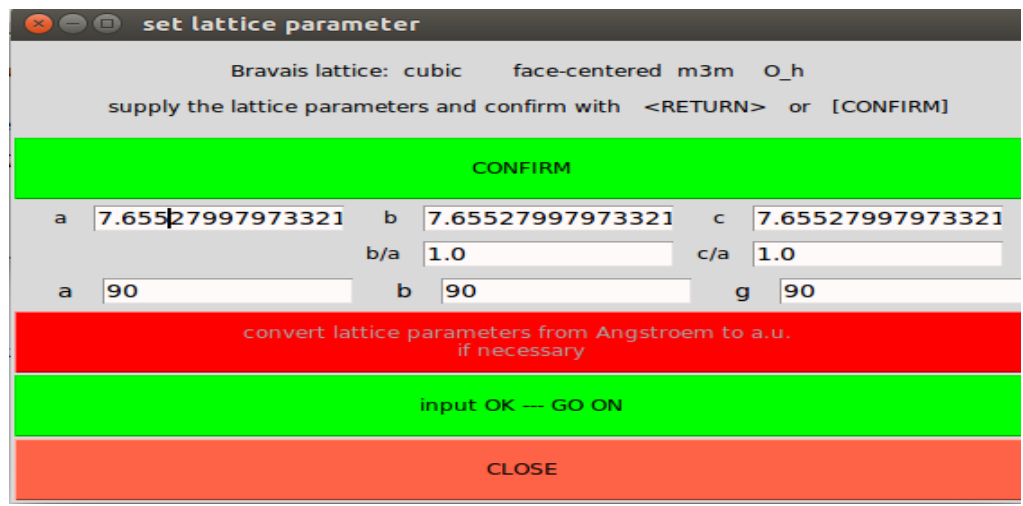

*Figure 5: This figure displays a dialog box titled "set lattice parameter" from the software application, related to materials science or crystallography.*

*Figure 6: Information on the structure of the Heusler Co2FeGe.*

And so, we obtain all the necessary information about the Co2FeGe compound to generate the AKAI-KKR input for the Go instruction calculation.

During the process of generating our Go Input, we choose the GGA (generalized gradient approximation) as the approximation, with a maximum of 520 iterations and 15 K-points.

```
#-----------------------------------------------
dos  data/FeGeCo2  fcc   7.6553,1.0,1.0,90.0,90.0,90.0,
0.001   2.2    sra    gga91    mag    init
update   15    520    0.020
    4
#-----------------------------------------------
#  name of type, number of components, rtr,
#  field, lmax, atomic number, concentration
#-----------------------------------------------
 Fe    1    0    0.0    2    26         100
 Ge    1    0    0.0    2    32         100
 Co_1  1    0    0.0    2    27         100
 Co_2  1    0    0.0    2    27         100
    4
     0.00000000      0.00000000      0.00000000    Fe
     0.50000000      0.50000000      0.50000000    Ge
     0.25000000      0.25000000      0.25000000    Co_1
     0.75000000      0.75000000      0.75000000    Co_2
```

*Figure 7 The input for the Go calculation*